\documentclass[aps,prl,twocolumn,superscriptaddress,showpacs,amsmath,amssymb]{revtex4-1}

\usepackage[pdftex]{graphicx}

\newcommand{\Fref}[1]{Figure~\ref{#1}}
\newcommand{\fref}[1]{Fig.~\ref{#1}}

\begin{document}

\title{Doping - dependent anisotropic superconducting gap in Na$_{1-\delta}$(Fe$_{1-x}$Co$_x$)As}

\author{K.~Cho}
\affiliation{The Ames Laboratory, Ames, IA 50011, USA}
\affiliation{Department of Physics and Astronomy, Iowa State University, Ames, IA 50011, USA}

\author{M.~A.~Tanatar}
\affiliation{The Ames Laboratory, Ames, IA 50011, USA}
\affiliation{Department of Physics and Astronomy, Iowa State University, Ames, IA 50011, USA}

\author{N.~Spyrison}
\affiliation{The Ames Laboratory, Ames, IA 50011, USA}
\affiliation{Department of Physics and Astronomy, Iowa State University, Ames, IA 50011, USA}

\author{H.~Kim}
\affiliation{The Ames Laboratory, Ames, IA 50011, USA}
\affiliation{Department of Physics and Astronomy, Iowa State University, Ames, IA 50011, USA}

\author{G.~Tan}
\affiliation{Department of Physics and Astronomy, The University of Tennessee, Knoxville, TN 37996, USA}
\affiliation{College of Nuclear Science and Technology, Beijing Normal University, Beijing 100875, China}

\author{P.~Dai}
\affiliation{Department of Physics and Astronomy, The University of Tennessee, Knoxville, TN 37996, USA}

\author{C.~L.~Zhang}
\affiliation{Department of Physics and Astronomy, The University of Tennessee, Knoxville, TN 37996, USA}

\author{R.~Prozorov}
\email[Corresponding author: ]{prozorov@ameslab.gov}
\affiliation{The Ames Laboratory, Ames, IA 50011, USA}
\affiliation{Department of Physics and Astronomy, Iowa State University, Ames, IA 50011, USA}

\date{\today}

\begin{abstract}
The London penetration depth was measured in single crystals of self-doped Na$_{1-\delta}$FeAs (from under - to optimal - doping, $T_c$ from 14 to 27~K) and electron-doped Na(Fe$_{1-x}$Co$_{x}$)As with $x$ ranging from undoped, $x=0$ to overdoped $x= 0.1$. In all samples, the low - temperature variation of the penetration depth exhibits a power-law dependence, $\Delta \lambda (T) = AT^n$, with the exponent that varies in a dome - like fashion from $n \sim 1.1$ in the underdoped, reaching a maximum of $n \sim 1.9$ in the optimally doped and decreasing again to $n \sim 1.3$ on the overdoped side. While the anisotropy of the gap structure follows universal dome-like evolution, the exponent at the optimal doping, $n \sim 1.9$, is lower than in other charge - doped Fe-based superconductors (FeSCs). The full - temperature range superfluid density, $\rho_s(T) = \left(\lambda(0)/\lambda(T)\right)^2$, at the optimal doping is also distinctly different from other charge - doped FeSCs but is similar to isovalently - substituted BaFe$_2$(As$_{1-x}$P$_x$)$_2$, believed to be a nodal pnictide at the optimal doping. These results suggest that the superconducting gap in Na(Fe$_{1-x}$Co$_{x}$)As is highly anisotropic even at the optimal doping.
\end{abstract}



\maketitle
The Fe-based superconductors (FeSCs) represent a rich variety of materials \cite{Mazin2010Nature,Paglione2010review,Chubukov2012ARCMP,Hirschfeld2011ROPP, Prozorov2011RPP}. Despite this chemical diversity, a generic trend is that a superconducting ``dome" in the temperature versus doping phase diagram forms in the vicinity or coexisting with the long range magnetic order. Instability of electronic system caused by the proximity of magnetism and superconductivity is considered as a key to understand pairing mechanism in the FeSCs \cite{Mazin2010Nature}. Contrary to the cuprate high-temperature superconductors, which show robust nodal $d$-wave superconducting gap for all doping types and levels, the gap structure of FeSCs is strongly doping-dependent varying from isotropic to nodal gap structure \cite{Gordon2009PRB, Martin2010FeNinodes, Tanatar2010PRL, Prozorov2011RPP, Reid2010PRB,Chubukov2012ARCMP, Hirschfeld2011ROPP}. An important question at this moment is whether there is any universality in doping dependence of the gaps among different FeSC materials.

The universal trend was suggested in studies of 122 family. In particular, the gap structure of both electron-doped Ba(Fe$_{1-x}$T$_x$)$_2$As$_2$ (``BaT122", T = transition metal) \cite{Gordon2009PRB, Martin2010FeNinodes, Tanatar2010PRL, Reid2010PRB, Ren2011arXiv, Budko2009PRB_SpecificHeat, MuTanigaki2011PRB} and hole-doped (Ba$_{1-x}$K$_x$)Fe$_2$As$_2$ (``BaK122") \cite{Kim2011arXiv, Reid2011arXiv, Dong2010PRL, Hashimoto2010PRB_KFe2As2} are nodeless and isotropic at the optimal doping and become highly anisotropic (perhaps nodal) at the dome edges. The isovalent-doped LiFe(As,P) \cite{Kim2011PRB, Hashimoto2012PRL} and layered Ca$_{10}$(Pt$_3$As$_8$)[(Fe$_{1-x}$Pt$_x$)$_2$As$_2$]$_5$ (``Ca10-3-8") compounds \cite{Cho2012PRB} obey this trend as well.
The only exception to this universal trend is found in isovalent - doped materials such as BaFe$_2$(As$_{1-x}$P$_x$)$_2$ (``BaP122") \cite{Hashimoto2010PRB, Yamashita2009PRB} and Ba(Fe$_{1-x}$Ru$_x$)$_2$As$_2$ \cite{Qiu2011arXiv} where nodal behavior was concluded even at the optimal doping. It was suggested by Kuroki \emph{et al.} that nodal vs. nodeless behavior is controlled by the pnictogen height at least in a 1111 family \cite{Kuroki2009PRB} and Hashimoto \emph{et al.} extended this suggestion to 111 and 122 families \cite{Hashimoto2012PRL}. It is also believed that the same parameter controls the transition temperature, $T_c$ \cite{Lee2008JPSJ, Mizuguchi2010SST}. Other factors that might control nodal vs. nodeless gap are doping-dependent electronic band structure and competition between intraband and interband interactions, taking into account different orbital contents in each band \cite{Chubukov2012ARCMP, Hirschfeld2011ROPP, Thomale2011PRL}. Therefore, it is important to study other Fe-based systems to test these ideas and determine possible universal trends, if they exist.

NaFeAs is one of few stoichiometric FeSCs with $T_c \sim$ 12 K \cite{Chu2009PC}, isostructural with LiFeAs ($T_c \sim$ 18 K). Application of 3 GPa pressure raises $T_c$ to 31 K, indicating its effective-underdoped nature \cite{Zhang2009EPL_NaFeAs} distinct from the slightly overdoped nature of stoichiometric LiFeAs \cite{Gooch2009EPL}. Supporting this underdoped nature of parent NaFeAs, coexistence of magnetism and superconductivity was found in NMR, neutron scattering, electronic transport and $\mu$SR measurements \cite{Kitagawa2010JPSJ, Parker2010PRL, Chen2009PRL, Parker2009CC}. The coexistence region extends to optimal-doping in NaFeAs with electron Co-substitution Fe,\cite{Parker2010PRL}. The angle-resolved photoemission spectroscopy (ARPES) study in near optimally-doped NaFe$_{0.95}$Co$_{0.05}$As revealed the quasi-nested Fermi surfaces connected by $Q$ = ($\pi$, $\pi$) wave vector, suggesting an important role of magnetic instability in superconductivity \cite{Liu2011PRB}, similar to the 122 family. This is in stark contrast to the absence of nesting in LiFeAs \cite{Borisenko2010PRL}, despite recent reports from inelastic neutron scattering, indicating incommensurate magnetic correlation in LiFeAs \cite{Qureshi2012PRL}. The gaps obtained in the ARPES study in NaFe$_{0.95}$Co$_{0.05}$As are isotropic \cite{Liu2011PRB}. However, they persist well above $T_c$ and thus are unlikely related to superconductivity. Scanning tunnelling spectroscopy revealed V - shaped curves \cite{YangWen2012arXiv}, which may be consistent with both nodal and nodeless gaps. Specific heat measurements revealed a jump at $T_c$, but did not extend to low enough temperature to reveal characteristic nodal vs. nodeless behavior \cite{WangChen2012PRB}. Considering the ambiguity of the phonon contribution subtraction and multi-gap nature of superconductivity in this compound, in the current stage the specific heat results cannot rule out the existence of (at least one) highly anisotropic gap. Therefore the important question on the anisotropy of superconducting gaps in NaFe$_{1-x}$Co$_{x}$As remains open.

In this work, we study doping-evolution of the superconducting gap structure in NaFeAs single crystals using a high resolution tunnel-diode resonator (TDR) technique. The doping level was controlled with either self-doping in Na$_{1-\delta}$FeAs caused by oxidative environmental deintercalation (underdoped to optimally doped range) or Co-doping in Na(Fe$_{1-x}$Co$_{x}$)As (spanning the whole superconducting dome). We found that, similar to most FeSCs, the variation of the London penetration depth is best described by a power-law function, $\Delta \lambda$ = $AT^n$, with the exponent $n$ and pre-factor $A$ varying systematically with doping in a dome-like fashion. The values of $n$ in the underdoped ($n\sim 1.1$) and heavily overdoped ($n\sim 1.3$) compositions strongly suggest the superconducting gap with line nodes. Even at the optimal doping the exponent $n \sim 1.9$ is notably lower than in the other charge - doped FeSCs \cite{Prozorov2011RPP}. The superfluid density, $\rho_s(T)$, at the optimal doping is also incompatible with isotropic full gap. Instead, $\rho_s(T)$ at $x =$ 0.05 is remarkably similar to that of optimally isovalent-doped BaFe$_2$(As$_{1-x}$P$_x$)$_2$, a known nodal pnictide superconductor even at the optimal doping \cite{Hashimoto2010PRB}. Our observations suggest that, despite the universal tendency for highest anisotropy at the dome edges in LiFe(As,P) \cite{Kim2011PRB, Hashimoto2012PRL}, BaCo122 \cite{Reid2010PRB}, BaK122 \cite{Kim2011arXiv} and a layered Ca10-3-8 system \cite{Cho2012PRB}, the gap structure at the dome center can highly anisotropic even in the charge-doped systems.

Single crystals of Na(Fe$_{1-x}$Co$_{x}$)As with $x =$ 0, 0.02, 0.025, 0.05, 0.08, and 0.10 were synthesized by sealing in the mixture of Na, Fe, As and Co together in Ta tubes and heating at 950 $^{\circ}$C, followed by 5 $^{\circ}$C/hour cooling down to 900 $^{\circ}$C \cite{He2010PRL}. We determined $x$ as used in load to crystal growth.
The samples were stored and transported in sealed containers filled with an inert gas. Sample preparation for TDR measurements was done quickly in air within about 5 minutes to minimize uncontrolled environmental exposure which can induce an increase of $T_c$ \cite{Todorov2010CM, Tanatar2012PRB}. The process of bulk ``self-doping'' can be well-controlled using reaction with Apiezon N-grease, and can be facilitated by ultrasonic treatment stimulating deintercalation of Na $^+$ ions. Duration of the sonication treatment was used to control $T_c$ of sample, see Ref.~\onlinecite{Tanatar2012PRB} for details. The low-temperature variation of the in-plane London penetration depth, $\Delta \lambda (T)$,  was measured using the TDR technique described elsewhere \cite{Prozorov2000PRB, Prozorov2006SST,Prozorov2011RPP}.

\begin{figure}[tb]
\includegraphics[width=7.5cm]{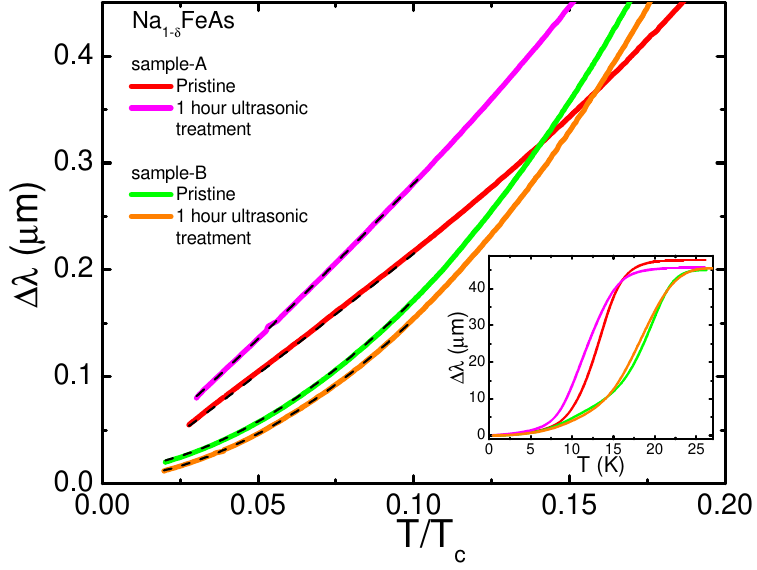}
\caption{(Color online) Low temperature part of $\Delta \lambda (T/T_c)$ in two Na$_{1-\delta}$FeAs samples (A and B) in pristine and ultrasonic treated (one-hour) states. Dashed lines are representative power-law fits up to $T/T_c = 0.1$ for quantitative analysis. Inset shows full-temperature range variation of $\Delta \lambda (T)$ for the same samples.}
\label{fig.1}
\end{figure}

\Fref{fig.1} shows the low temperature variation of $\Delta \lambda (T)$ in two samples A and B of Na$_{1-\delta}$FeAs crystals, which shows a clear evolution from almost $T$-linear (pristine) to almost quadratic (one-hour ultrasonic treatment) behavior. Measurements were done immediately after opening the ampoule (pristine state) and repeated after one-hour ultrasonic treatment, which increased $T_c$ to 22~K (See the inset). Successive treatments increased $T_c$ to a maximum value of $\sim$ 27 K (after 2 and 3 hour treatments), but longer treatments lead to sample degradation.

\begin{figure}[tb]
\includegraphics[width=7.5cm]{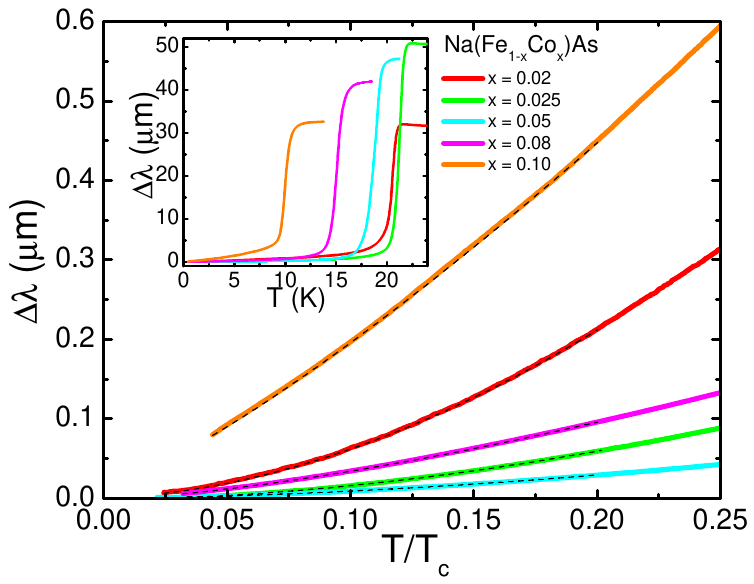}
\caption{(Color online) (a) Low-temperature part of $\Delta \lambda (T/T_c)$ in Na(Fe$_{1-x}$Co$_{x}$)As for $x =$ 0.02, 0.025, 0.05, 0.08, and 0.10. Dashed lines are representative power-law fits up to $T/T_c = 0.2$  for quantitative analysis. Fitting results are summarized in \fref{fig.3}. Inset shows full-temperature range variation of $\Delta \lambda (T)$ for the same samples.}
\label{fig.2}
\end{figure}

\Fref{fig.2} shows the low temperature variation of $\Delta \lambda (T)$ in Na(Fe$_{1-x}$Co$_{x}$)As for $x =$ 0.02, 0.025, 0.05, 0.08, and 0.1. Judging by $T_c(x)$ in the inset, the samples cover the superconducting dome from lightly underdoped to heavily overdoped regime. For a quantitative analysis of low temperature behavior, a fit to power-law function, $\Delta \lambda (T) = A T^n$, was performed over a temperature interval from the base temperature up to  several upper limits, varying from $T_c$/10 to $T_c$/5. The representative fits for $T_c$/5 limit are shown as dashed lines.

\begin{figure}[tb]
\includegraphics[width=7cm]{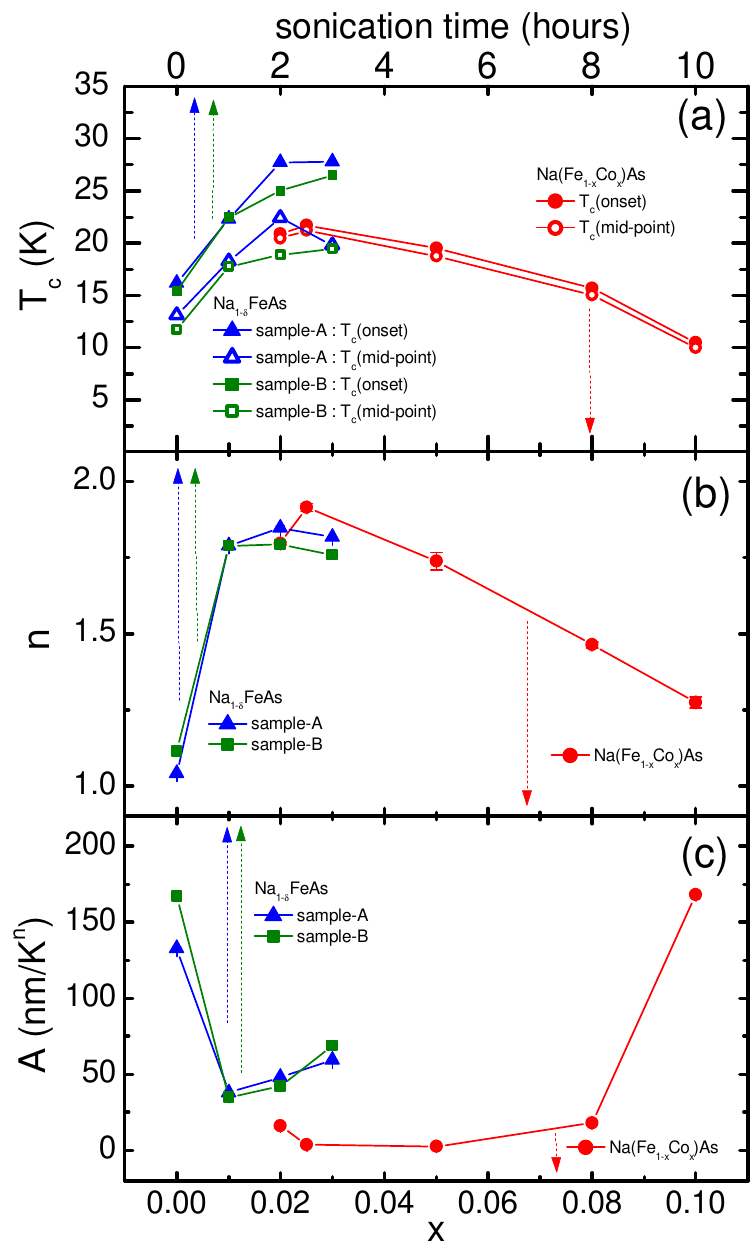}
\caption{ (Color online) Phase diagram of NaFeAs system. (a) superconducting ``dome", $T_c(x)$, (b) power-law exponent $n(x)$, and (c) prefactor $A(x)$. Self-doped results were placed against sonication time (top-axis) and scaled by their $T_c$ (see text). Circles show the data for Co-doped samples. Other symbols show the data for self-doped samples.}
\label{fig.3}
\end{figure}

\Fref{fig.3} presents a summary of our results for NaFeAs system. A doping dependence of $T_c$ (panel a) in Na(Fe$_{1-x}$Co$_{x}$)As is shown against actual $x$ (bottom axis). The data for Na$_{1-\delta}$FeAs are plotted against sonication time (top axis) after scaling that the maximum $T_c$ is matched with that from the Co-doped case (i.e., 2.5-hour sonication treatment equivalent to $x =$ 0.025). The parameters of power-law fit, $\Delta \lambda (T)=AT^n$, are summarized in \fref{fig.3} (b) and (c). For each data set, fitting was performed for several temperature ranges to examine the robustness of the fitting parameters. In Na$_{1-\delta}$FeAs, the exponent $n$ is about 1.1 in the most underdoped composition (pristine samples) and increases to 1.8 after 3 hour sonication when the highest $T_c \approx$ 27~K is reached. Similarly, larger exponent, $n \sim$ 1.9, was obtained in Na(Fe$_{1-x}$Co$_{x}$)As for $x =$ 0.025 (optimal doping) and it monotonically decreased to $n \sim$ 1.3 for overdoped $x =$ 0.10. In general, the results for both Na$_{1-\delta}$FeAs and Na(Fe$_{1-x}$Co$_{x}$)As follow universal trend that the gap anisotropies are highest at the dome edges and smallest at the dome center. The values at the dome edges are notably below the lowest possible value ($n \sim 1.7$) expected from $ s_{\pm}$ scenario with strong pair-breaking scattering \cite{Prozorov2011RPP}, suggesting the formation of nodes. In addition, the low value of exponent $n \sim 1.9$ at the dome center makes a clear distinction of Na$_{1-\delta}$(Fe$_{1-x}$Co$_{x}$)As from the other charge - doped FeSCs. The exponents close to $n \sim 2$ can be explained by strong pair breaking effect in $s_{\pm}$ or $d$-wave pairing scenarios.

\begin{figure}[tb]
\includegraphics[width=8cm]{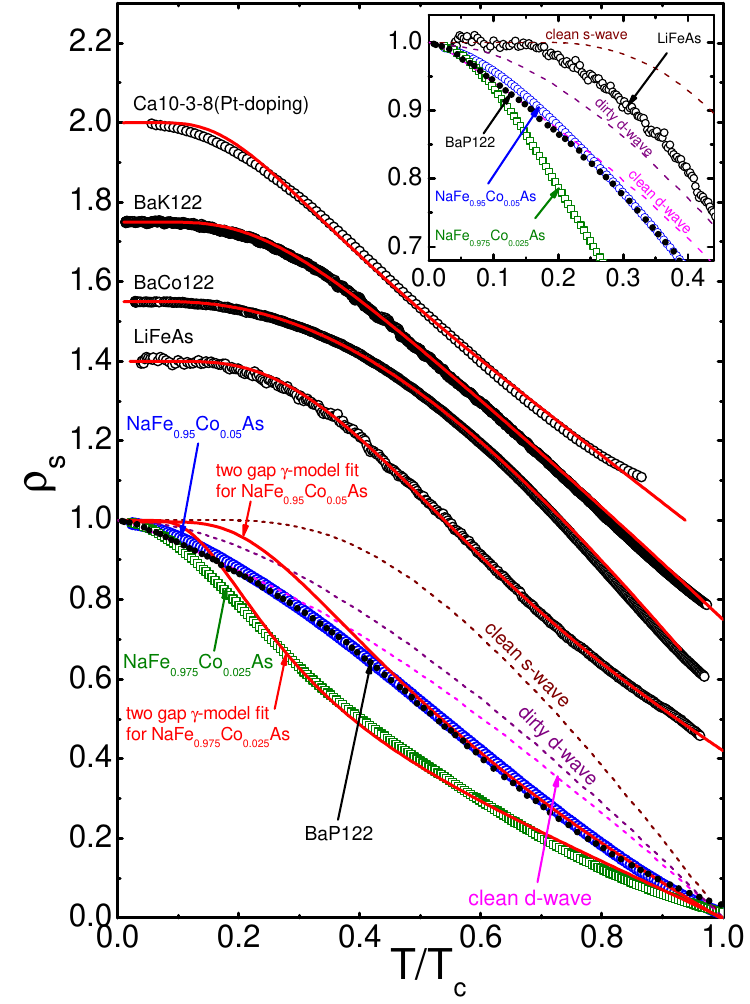}
\caption{ (Color online) Experimental superfluid densities, $\rho _{s} (T)$, of optimally-doped FeSCs. Top four curves for LiFeAs \cite{Kim2011PRB}, BaCo122 \cite{Prozorov2011RPP}, BaK122 \cite{Kim2011arXiv}, and Ca10-3-8 \cite{Cho2012PRB} were offset for clarity. Red solid curves show the best fits to an isotropic s-wave two-gap $\gamma$-model \cite{Kogan2009gamma} demonstrating good agreement with the data. In a clear contrast, the $\rho_{s}$ for optimally - doped NaFe$_{0.975}$Co$_{0.025}$As (green open-squares) and NaFe$_{0.95}$Co$_{0.05}$As (blue open-circles) cannot be fitted with this fully gapped model, especially at the lowest temperatures, zoomed at in the inset. For comparison, we also show the data for optimally - doped BaP122 ($x=0.33$, black solid dots) \cite{Hashimoto2010PRB} as well as single $d-$ and $s-$ gaps (dashed lines) in both clean and dirty limits.}
\label{fig.4}
\end{figure}

For a more complete analysis of superconducting gap structure, especially in a multi-gap system, the low-temperature power-law behavior is not sufficient, as it only reflects quasiparticle excitations around gap minima. \Fref{fig.4} shows normalized superfluid density, $\rho_{s} (T)=\left(\lambda(0)/\lambda(T) \right)^2$, plotted against $T/T_c$ for several optimally doped FeSCs. To calculate $\rho_{s}$ in NaFe$_{0.975}$Co$_{0.025}$As (green open-squares) and NaFe$_{0.95}$Co$_{0.05}$As (blue open-circles), we used the values of $\lambda _{0}$ = 410 nm and 354 nm, respectively, obtained from $\mu$SR measurements \cite{Parker2010PRL}. The $\rho_{s}$ for other materials (offset vertically  for clarity in \Fref{fig.4}) include LiFeAs \cite{Kim2011PRB}, BaCo122 \cite{Prozorov2011RPP}, BaK122 \cite{Kim2011arXiv}, and Ca10-3-8 \cite{Cho2012PRB}, all measured in our group. Red solid curves are the best fits to a self-consistent isotropic s-wave two-gap $\gamma$-model \cite{Kogan2009gamma} showing good fit quality in all four cases. In stark contrast, an attempt to fit the data for NaFe$_{0.975}$Co$_{0.025}$As and NaFe$_{0.95}$Co$_{0.05}$As (both are near the optimal doping) with the same $\gamma-$ model clearly fails especially at the low-temperature end,  where the distinction between full and nodal gaps is most pronounced, as zoomed in the inset of \fref{fig.4}. For comparison, we also show $\rho_s(T/T_c)$ expected for a single-gap $s$- and $d$-wave pairings both in clean and dirty limits (dashed curves). These single-gap curves clearly deviate in the whole temperature range, confirming multi-gap nature of superconductivity in Na(Fe$_{1-x}$Co$_{x}$)As (as well as other FeSCs). Surprisingly, $\rho_s(T/T_c)$ of NaFe$_{0.95}$Co$_{0.05}$As is virtually the same as that of optimally - doped BaP122 ($x$ = 0.33, black solid dots), a known nodal isovalent substituted pnictide \cite{Hashimoto2010PRB}, and shows notable difference from the exponential saturation in a fully gapped LiFeAs \cite{Kim2011PRB,Hashimoto2012PRL}.

Our results show that the superconducting gap even at the optimal doping is highly anisotropic in charge-doped Na(Fe$_{1-x}$Co$_{x}$)As. Similar to other FeSCs it becomes more anisotropic upon departure from the optimal doping. To explain such sensitivity to the doping level and appearance of nodes in some materials, several proposals have been put forward. (i) Following a suggestion by Kuroki \emph{et al.} that nodal vs. nodeless behavior is controlled by pnictogen height in 1111 family \cite{Kuroki2009PRB}, Hashimoto \emph{et al.} extended this suggestion to 111 and 122 families \cite{Hashimoto2012PRL}. (ii) Nodal structure of the superconducting gap was linked with the evolution of the Fermi surface topology and orbital content in different bands \cite{Chubukov2012ARCMP, Hirschfeld2011ROPP, Thomale2011PRL_2, Thomale2011PRL, NicholsonDagotto2011PRL}. The bandstructure of NaFeAs is significantly different from 122 compounds, which itself varies significantly between different hole, electron and isovalent substitutions with no apparent correlation that may explain appearance of nodes only in some compounds. (iii) Another approach is to incorporate changes in both, the bandstructure and renormalization of intra-band and in-band Coulomb interactions. According to Ref. \onlinecite{Chubukov2012ARCMP}: (1) for weakly and moderately electron-doping, the propensity of $s_{\pm}$ and $d-$wave pairing channels are comparable; (2) weakly and moderately hole-doping, $s_{\pm}$ pairing is dominant; and (3) for strongly electron- and hole-doping, $d$-wave channel is dominant. From this classification, regime 1 seems to be most relevant for the optimally-doped Na(Fe$_{1-x}$Co$_{x}$)As and regime 3 is appropriate for the overdoped compositions, consistent with our results.

In conclusion, a combined study of self-doped Na$_{1-\delta}$FeAs and charge-doped Na(Fe$_{1-x}$Co$_{x}$)As establishes the systematics of the superconducting gap evolution with doping in the 111 compounds. Similar to other FeSCs, we find a universal tendency of developing highest anisotropy at the dome edges. However, we show that, even at the optimal doping, the low temperature penetration depth study indicates highly anisotropic superconducting gap. In addition, temperature - dependent superfluid density at the optimal doping is remarkably similar to $\rho_s(T)$ of isovalent-doped BaFe$_2$(As$_{1-x}$P$_x$)$_2$, a known nodal pnictide superconductor at the optimal doping.

After posting the preprint of this paper \cite{Cho2012arXiv_NaFeCoAs}, two studies using crystals from Prof. X.~H.~Chen group \cite{YangWen2012arXiv, WangChen2012PRB} suggested full isotropic superconducting gap in NaFe$_{1-x}$Co$_x$As at optimal Co-doping. To check if the difference in the conclusions is caused by the difference in the samples of this reactive material, we asked Prof. X.~H.~Chen to send us crystals for a cross - examination. We received and measured optimally doped samples with $x$=0.028. The results were virtually identical (within the noise level) to those presented in Fig.~\ref{fig.3} and~\ref{fig.4}, ruling out the sample dependence of our conclusions. Possible reasons for the discrepancy with other studies are discussed in the introduction.

\begin{acknowledgments}
We thank A. Chubukov, P. Hirschfeld, R. Thomale, A. Kaminski, Hai-Hu Wen, X. H. Chen and Shi-Yan Li for useful discussions. We specially thank X. H. Chen and X. G. Luo for supplying crystals for a cross - examination study. Work at the Ames Laboratory was supported by the U.S. Department of Energy, Office of Basic Energy Sciences, Division of Materials Sciences and Engineering under contract No. DE-AC02-07CH11358.. The work at University of Tennessee was supported by U.S. DOE BES under Grant No. DE-FG02-05ER46202 (P.D.).
\end{acknowledgments}

\end{document}